\documentclass[prd,preprint,showpacs,preprintnumbers,amsmath,amssymb]{revtex4}
\usepackage{dcolumn}% Align table columns on decimal point
\usepackage{bm}% bold math
\def\cC{{\cal C}}
\def\cM{{\cal M}}
\def\cP{{\cal P}}
\def\ex{{\rm exp}}
\def\cA{{\cal A}} 

\def\cL{{\cal L}}
\def\cS{{\cal S}}

\def\im{\sqrt{-1}}
\def\rd{{\rm d}}
\def\Tr{{\rm Tr}}
\def\Met{{\rm Met}}
\def\Diff{{\rm Diff}}
\def\Weyl{{\rm Weyl}}
\def\Cyl{{\rm Cyl}}
\def\Disc{{\rm Disc}}
\def\Hol{{\rm Hol}}
\def\CKD{{\rm CKD}}

\begin{document}

%\preprint{APS/123-QED}

\title{Strings with extended non-Abelian gauge interaction}% Force line breaks with \\

\author{Zden\v{e}k Kopeck\'{y}}

\email{kopecky@physics.muni.cz}

\affiliation{Institute for Theoretical Physics and Astrophysics\\ Masaryk University\\Kotl\'{a}\v{r}sk\'{a} 2, 611 37 Brno\\ Czech Republic}


\begin{abstract}
The new generalization of the gauge interaction for the bosonic strings is found.
We consider some quasiequivariant maps from the space of metrics on the worldsheet to the space of $n$-tuples of one- and two-dimensional loops.
The two-dimensional case is based on the cylinders interacted with a path space connection.
The special 2-gauge string model is formulated using two 1-connections, non-Abelian background symmetric tensor field and non-Abelian 2-form.
The branched non-Abelian space-time is the result of our construction.
\end{abstract}

\maketitle
\section{Introduction}
Gauge interaction in the open string theory was introduced by adding the Wilson loops of a connection 1-form along each component of the boundary of the string worldsheet. A two-dimensional generalization (2-gauge string theory) of the previous one-dimensional construction is not known. Many motivations come from the formulation of the gauge theory on the path space and especially from the geometry of the non-Abelian Stokes theorem \cite{Polyakov,FNep,Nep,MenskyBook,AlvST,CatST}. The two-dimensional surface is considered as an ordered one, in other words, covered by a path in the space of paths on the surface. A no-go theorem of Teitelboim \cite{Teitel} states that 2-dimensional non-Abelian gauge interaction is not compatible with  the reparametrization invariance.

The works \cite{MenskyStr,Kapust,Kalk,Chep,Hof,Pfeif1,Pfeif2,Lah,Schr,Jurco} contain some contributions to
the 2-gauge string theory building. Their approaches are based on the path space or the gerb geometry.

In this paper the 2-Wilson loops have been incorporated directly into the Polyakov path integral.
%%%%%1D%%%%%%
\section{String worldsheets with moving Wilson loops}
Let $\Sigma$ be an oriented string worldsheet modeled by an open or closed orientable compact surface and we denote by ${\rm CTop}_{2}^{+}$ nonequivalent topologies of such surfaces. Let $\Met(\Sigma)$ be the space of Eucledean metrics on the surface $\Sigma$. The string worldsheet is embedded into a target space-time manifold $M$:
\begin{equation} 
X: \Sigma \rightarrow M.
\end{equation}
Let the  manifold $M$ be the base of a gauge bundle with a compact gauge group and a connection $A$. There is an  $n$-tuple of loops $\cC=(C_{1},\ldots,C_{n}) \in (\cL\Sigma)^n$ embedded on the surface where $\cL\Sigma$ denotes the loop space: $\cL\Sigma=C^{\infty}(S^{1},\Sigma)$.
Our aim is to modify the string partition function by Wilson loops of the connection $A$ around the $n$-tuple $\cC$:
\begin{equation}
 W_{F}^{(1)}[\cC]=\int D\gamma\,DX \exp (-S(\gamma,X)) \prod_{C_{i} \in \cC} \Tr P \exp\int_{C_{i}}A,
 \label{Wil0}
\end{equation}
where $S(\gamma,X)$ is the string action in the background $(G,B)$ of metric tensor and antisymmetric form, respectively:
\begin{equation}
 S(\gamma,X)=\frac{1}{2}\int_{\Sigma} (G_{\mu\nu}\rd X^{\mu} \wedge \ast \rd X^{\nu}-\im \, B_{\mu\nu}\rd X^{\mu} \wedge \rd X^{\nu}).
 \label{Sac}
\end{equation}
Let the positions of the loops in (\ref{Wil0}) be a general ones not fixed on the boundaries of the worldsheet, as one
usually uses in string theory. A dependence of the system loops $\cC$ on the surface metric is considered:
\begin{equation}
 \Met(\Sigma)\ni \gamma \rightarrow \cC(\gamma)=(C_{1}(\gamma),\ldots,C_{n}(\gamma)) \in (\cL \Sigma)^{n}.
 \label{MetF}
\end{equation}
The distribution $\cC(\gamma)$ of $n$-tuples over the space $\Met(\Sigma)$ in (\ref{MetF}) cannot be arbitrary.
We impose the quasiequivariant condition on the structure function (\ref{MetF}) to save the Weyl-diffeomorphism symmetry:
\begin{eqnarray}
&&\forall \gamma \in \Met(\Sigma),\, \forall  w=(f,e^{2\phi}) \in \Diff^{+}(\Sigma) \ltimes \Weyl (\Sigma), \label{ConDI} \\
&&\exists \, \tilde{f_K}(\gamma) \in \CKD(\Sigma,\gamma),\,\,\exists \, \tilde{F}^{S^{1}}(\gamma)\in (\Diff^{+}(S^{1}))^{n},\nonumber\\
&&\exists \, \tilde{\Pi}(\gamma)\in S_{n}:\,\, \cC(w(\gamma))=(f\tilde{f}_{K}(\gamma))^{-1}[\cC(\gamma)\circ \tilde{F}^{S^{1}}(\gamma)]_{\tilde{\Pi}(\gamma)}, \nonumber
\end{eqnarray}
where $S_{n} \ni \Pi \mapsto [\ldots]_{\Pi}$ denotes the action of permutation group $S_{n}$ on the loops, $\CKD(\Sigma,\gamma)$ is the group of conformal Killing diffeomorphisms of the metric $\gamma$ and $\Weyl(\Sigma)$ is the group of Weyl transformations.
The integration over all metrics in (\ref{Wil0}) substantially depends on the system function $\cC$ (\ref{MetF}). The property (\ref{ConDI}) allows us to restrict the integration on the standard moduli space:
\begin{eqnarray}
 \cM(\Sigma)=\frac{\Met(\Sigma)} {\Diff^{+}(\Sigma)\ltimes \Weyl(\Sigma)}\,.
 \label{Mod0}
\end{eqnarray}
The extended moduli space is introduced
\begin{equation}
 \cM^{C}_{n}(\Sigma)=\frac{\Met(\Sigma)\times (\cL \Sigma)^{n}} {\thicksim_{C}},
 \label{ExMod} 
\end{equation}
where $\thicksim_{C}$ is the relation of equivalence on the space $\Met(\Sigma)\times (\cL \Sigma)^{n}$:
\begin{widetext}
\begin{eqnarray}
 &&\thicksim_{C}:\,\, (\gamma,\cC)\cong (\gamma',\cC') \Leftrightarrow
 \gamma'=w(\gamma)\,\,{\rm and}\,\, \cC'=f[\cC \circ F^{S^{1}}]_{\Pi}, \label{DistF}\\\nonumber
 &&  w=(f,e^{2\phi}) \in \Diff^{+}(\Sigma) \ltimes  \Weyl (\Sigma),\,
 F^{S^{1}} \in (\Diff^{+}(S^{1}))^{n}, \, \Pi \in S_{n}.
 \end{eqnarray}
 \end{widetext}
The extended moduli space (\ref{ExMod}) can be endowed with the canonical projection to the standard moduli space:
\begin{equation}
 \cM^{C}_{n}(\Sigma)
 \rightarrow
 \cM(\Sigma).
 \label{LFB}
\end{equation}
The bundle (\ref{LFB}) we call the structure bundle.
Let $\Gamma (\cM^{C}_{n}(\Sigma) \rightarrow \cM(\Sigma))$ be the space of sections of the structure bundle (\ref{LFB}).
%We indetify a $n$-tuples with the property (\ref{ConDI}) and sections of (\ref{LFB}).
Two loop systems $\cC$ and $\cC^{'}$ with the property (\ref{ConDI}) define a section of the bundle if and if
 \begin{equation}
  \forall \gamma\in\Met(\Sigma): \,\, \cC^{'}(\gamma)=(\tilde{f}_{K}(\gamma))^{-1}[\cC(\gamma)\circ
  \tilde{F}^{S^{1}}(\gamma)]_{\tilde{\Pi}(\gamma)},
  \label{SecCond}
 \end{equation}
where $\tilde{F}^{S^{1}}(\gamma) \in \Diff^{+}(S^{1}),\,\,\tilde{\Pi}(\gamma) \in S_{n}$ and $\tilde{f}_{K}(\gamma)\in \CKD(\Sigma,\gamma)$.

The resulting path integral will depend on a sequence of extended bundle sections sets. The sequence is parametrized  by the topology discrete parameter $g \in {\rm CTop}^{+}_{2}$:
\begin{equation}
 \cS_{g} =\{ s_{m} \in \Gamma (\cM^{C}_{n_{m}}(\Sigma_{g}) \rightarrow \cM(\Sigma_{g})),\,\, m \in {\bf N}, n_{m} \in 
 {\bf N_0}\}.
\end{equation}
Now we are able to write explicitly the partition function
\begin{widetext}
\begin{equation}
 W^{(1)}_{F}[\{\cS_{g}\}]=\sum_{g} \sum_{s \in \cS_{g}} \int_{\cM_{g}} [D\gamma] \int DX \exp (-S(\gamma,X)) \prod_{C_{i}(\gamma) \in  \cC(\gamma)}
  {\Tr P \exp\int_{C_{i}(\gamma)}A},
 \label{FE1}
\end{equation}
\end{widetext}
where $\cC(\gamma)$ is a representative for the element $s\in \cS_{g}$ and $[D\gamma]$ is the integral over the moduli space $\cM_{g}=\cM(\Sigma_{g})$.

The sum over closed surfaces with trivial non-intersected loops provides an example of Eq. ({\ref{FE1}) for $n_{m}=m$.
%%%%%D2%%%%
\section{String worldsheets with 2-gauge interaction}
In this section we apply the method of moving loops to the case of two-dimensional cylinders on the worldsheet.

A proposal for 2-gauge string model in the context of the path group approach was given in \cite{MenskyStr}. 
Further applications of differential geometry on the path space are contained in \cite{AlvST,CatST}.

\subsection{String interaction with path space connection}
We denote by $\cP Y$ the space of smooth paths on a manifold $Y$. The space of smooth loops on $Y$ will be denoted by
$\cL Y$ and  $\cL \cP Y=C^{\infty}(S^{1}\times\,[0,1] ,Y)$ is  the space of cylinders on $Y$. Let a principal $G$-bundle $P_{G}=P(\cP M,G)$ be built on the space $\cP M$ with the compact structure gauge group $G$ and the bundle geometry is governed by a $G$-connection $\cA$. An embedding $X:~ \Sigma \rightarrow M$ from the string worldsheet to the target space $M$ induces the following map in the path spaces:
\begin{equation} 
 \widetilde{X}:~ \cP \Sigma \rightarrow \cP M,\, \cP\Sigma \ni p \rightarrow X\circ p.
\end{equation}
We pull back the $G$-bundle $P_G$ and consequently the connection $\cA$ using the map $\widetilde{X}$ on the space $\cP\Sigma$. The holonomy of the connection $\cA$ along a loop in the space $\cP\Sigma$  represented by a cylinder $\Cyl \in \cL\cP\Sigma$
\begin{equation}
 P ~\ex  \int_{\Cyl} \cA
 \label{Exp2}
\end{equation}
will be the basic geometry object in our construction. The path-ordered exponential in Eq. ({\ref{Exp2}) is taken with respect to the $S^{1}$ argument of the cylinder $\Cyl$.
Let $n$-tuples of cylinders be distributed over $\Met(\Sigma)$:
\begin{equation}
\Met(\Sigma)\ni \gamma \rightarrow \cC^{2}(\gamma)=(\Cyl_{1}(\gamma),\ldots,\Cyl_{n}(\gamma)) \in (\cL \cP \Sigma)^{n}.
 \label{StrC2}
\end{equation}
The string path integral will be modified in a similar way as in the one-dimensional case by the product of traces of holonomies connected with the $n$-tuple:
\begin{equation}
 W^{(2)}[\cC^{2}(\gamma)]=\prod_{i=1}^{n} \Tr P ~\ex  \int_{\Cyl_{i}(\gamma)} \cA.
 \label{W2Hol}
\end{equation}
The distribution (\ref{StrC2}) cannot be arbitrary to reduce the $\gamma$-integration in the path integral to the moduli space (\ref{Mod0}), therefore, we restrict (\ref{StrC2}) by the quasiequivariant condition
\begin{widetext}
\begin{eqnarray}
 &&\forall \gamma \in \Met(\Sigma),\, \forall w=(f,e^{2\phi}) \in \Diff^{+}(\Sigma) \ltimes \Weyl (\Sigma),
 \label{ConDI2}\\\nonumber
 &&\exists \, \tilde{f_K}(\gamma) \in \CKD(\Sigma,\gamma),\,\,\exists \, \tilde{F}(\gamma) \in
 (\Diff[\Cyl])^{n},\\\nonumber && \exists \, \tilde{\Pi}(\gamma)\in S_{n}:\,\,
 \cC^{2}(w(\gamma))=(f\tilde{f}_{K}(\gamma))^{-1}[\cC^{2}(\gamma)\circ \tilde{F}(\gamma)]_{\tilde{\Pi}(\gamma)},
 \end{eqnarray}
 \end{widetext}
where $\Diff[\Cyl]$ is given for a general connection:
\begin{equation}
 \Diff[\Cyl]=\Diff^{+}(S^{1})\times Id,
 \label{DiffInv} 
\end{equation} 
and in the particular case of reparametrization invariant connection:
\begin{equation}
 \Diff[\Cyl]=\Diff^{+}(S^{1}) \times \Diff^{+}([0,1]).
 \label{DiffInv2} 
\end{equation}  
 The extended moduli space which corresponds to the modification of the string path integral by the term (\ref{W2Hol}) is
\begin{equation}
 \cM^{C^{2}}_{n}(\Sigma)=\frac{\Met(\Sigma)\times (\cL\cP \Sigma)^{n}} {\thicksim_{C^{2}}},
\label{ExMod2} 
\end{equation}
where the equivalence relation is given by
\begin{widetext}
\begin{eqnarray}
 &&\thicksim_{C^{2}}:\,\, (\gamma,\cC^{2})\cong (\gamma',\cC'^{2}) \Leftrightarrow
 \gamma'=w(\gamma)\,\,{\rm and}\,\, \cC'^{2}=f[\cC^{2} \circ F]_{\Pi}, \\\nonumber && w=(f,e^{2\phi}) \in \Diff^{+}(\Sigma) \ltimes 
 \Weyl(\Sigma),\,
 F \in (\Diff[\Cyl])^{n}, \, \Pi \in S_{n}.
  \label{DistF2}
\end{eqnarray}
\end{widetext}
The structure bundle of the extended moduli space (\ref{ExMod2}) is
\begin{equation}
 \cM^{C^{2}}_{n}(\Sigma)
 \rightarrow
 \cM(\Sigma).
 \label{ExtM2}
\end{equation}
Finally, the partition function for our strings with the cylinder Wilson loops reads
\begin{widetext}
\begin{equation}
 W_{F}^{(2)}[\{\cS_{g}\}]=\sum_{g} \sum_{s \in \cS_{g}}\int_{\cM_{g}} [D\gamma] \int DX \exp (-S(\gamma,X))
 \prod_{\Cyl_{i}(\gamma) \in \cC^{(2)}(\gamma)} \Tr P \exp\int_{\Cyl_{i}(\gamma)}\cA,
\end{equation}
where

\begin{equation}
 g \in {\rm CTop}_{2}^{+},\,\,\cS_{g} =\{ s_{m} \in \Gamma (\cM^{C^{2}}_{n_{m}}(\Sigma_{g}) \rightarrow \cM(\Sigma_{g})),\,\, m \in {\bf N},\, n_{m} \in {\bf N_0}\}
\label{Sg2} 
\end{equation}
\end{widetext}
and $\cC^{2}(\gamma)$ is a representative for the element $s \in \cS_{g}$.
\subsection {Special 2-gauge strings}
In this part a non-Abelian string interaction is formulated using geometrical objects connected with the bundle of horizontal paths \cite{CatST}.
The horizontal paths with respect to a connection $A$ on a given principal bundle $P$ create the principal $G$-bundle $\cP_{A}$ that is an example of principal $G$-bundle on the path space. A connection $\bar{A}$ and a non-Abelian 2-form $B$ define the special connection on the $G$-bundle $\cP_{A}$ \cite{CatST}.

Let $R$ be an irreducible representation of the compact simple gauge group $G$. The product of $R$ and its contragradient representation $\bar{R}$ can be expanded to the direct sum of irreducible representations $R_{i}$:
\begin{eqnarray}
 R\otimes\bar{R}={\sum_{i}} R_{i}.
 \label{RepProd}
\end{eqnarray} 
The group $G$ acts by the similarity transformation on the space of the complex $N\times N$ matrices $M_{N}(C)$ where $N\,=\,{\rm dim}_{C}\, R$. The invariant subspace of the Hermitian matrices
\begin{eqnarray}
 {\rm Herm}(N) \subset M_{N}(C)
\end{eqnarray}
can be written as the real sum of the irreducible subspaces
\begin{eqnarray}
 {\rm Herm}(N)= {\sum_{i}} V_{R_{i}},
\end{eqnarray}
 where $V_{R_{i}}$ is the representation space for the representation $R_{i}$ from the expansion (\ref{RepProd}).
Let $P_{CP}$ be a fixed principal $G$-bundle over $M$, and we generate from the bundle $P_{CP}$ the associate vector 
bundles
\begin{eqnarray}
  E_{V_{R_{i}}}= P_{CP}\times_{R_{i}}V_{R_{i}}.
\label{Ebundl}
\end{eqnarray}
Our aim is to generalize the action (\ref{Sac}) by changing the symmetric tensor $G_{\mu \nu}$ and the antisymmetric tensor $B_{\mu \nu}$  by the bundle (\ref{Ebundl}) valued objects, in other words, they will be sections of the bundles $T_{Sym}^{2}(M,E_{V_{R{i}}})$ and $\Omega^{2}(M,E_{V_{R_{i'}}})$, respectively.

We begin with putting together an embedded cylinder $\Cyl \in \cL P \Sigma$, bundle (\ref{Ebundl}) valued tensors $(G,B)$, and a pair of connections $(A,\, \bar{A})$ in the representation $R$. As an extension of the holonomy construction for the special connection \cite{CatST}, a gauge invariant quantity, which comprises all these objects, can be constructed:
\begin{widetext}
\begin{eqnarray}
 W^{(2)}[\Cyl]=\Tr P_{s} ~\ex \left[-\frac{1}{2} \int_{0}^{1} ds\, \Hol(\bar{A},s)^{-1}\, SH(G,B,A,s)\,\Hol(\bar{A},s)\right],
\label{HolCyl} 
\end{eqnarray}  
where
\begin{eqnarray}
 &&\Hol(\bar{A},s)= P_{\bar{\bullet}}~ \ex  \int_{\Cyl(\bar{\bullet},0)} \bar{A},\,\,\bar{\bullet}=[0,s], \label{HAcyl}\\
 &&SH(G,B,A,s)=\\
 &&\int_{0}^{1} dt\, \left[u_{A}(s,t)^{-1}\Cyl_{(s,t)}^{*}(G_{\mu\nu}\rd X^{\mu} \wedge \ast \rd X^{\nu}-
 \im B_{\mu\nu}\rd X^{\mu} \wedge \rd X^{\nu})\,u_{A}(s,t)\right],\nonumber\\
 &&\Cyl_{(s,t)}^{*}(\omega)= \Cyl^{*}(\omega)(\frac{\partial}{\partial s},\frac{\partial}{\partial t})(s,t),\,\omega \in \Omega^{2}(\Sigma),\\
 &&u_{A}(s,t)=P_{\bullet}~ \ex \int_{\Cyl(s,\bullet)} A, \,\,\bullet=[0,t],\,\,(s,t)\in S^{1}\times [0,1].
\end{eqnarray}
\end{widetext}
Eq.~(\ref{HolCyl}) is invariant with respect to the local gauge transformation generated by a function $\tilde{g}(x)\in G$
on $M$:
\begin{eqnarray}
 &A& \rightarrow  \tilde{g}A\tilde{g}^{-1}+d\tilde{g}\tilde{g}^{-1},\\
 &\bar{A}&\rightarrow \tilde{g}\bar{A}\tilde{g}^{-1}+d\tilde{g}\tilde{g}^{-1},\\
 &B&\rightarrow \tilde{g}B\tilde{g}^{-1},\\
 &G&\rightarrow \tilde{g}G\tilde{g}^{-1}.
\end{eqnarray}
%The parallel transport operator needs insertion of the transiton functions on the overlaping of the trivialisation %maps. For sake of simplicity, we %have written the formulas \ref(  without these functions.
The disc is the result of transforming the $\bar{A}$-boundary of a cylinder
\begin{equation}
 \Cyl(\bar{\bullet},0),\,\bar{\bullet}=[0,1]
 \label{CylAB}
\end{equation} 
into a point. In the disc case Eq.~(\ref{HolCyl}) is simplified to
\begin{eqnarray}
 W^{(2)}[\Disc]=\Tr P_{s} ~\ex \left[-\frac{1}{2}\int_{0}^{1} ds\ SH(G,B,A,s)\right].
\end{eqnarray}

The partition function can be directly formulated:
\begin{widetext}
\begin{equation}
 W_{F}^{(2)}[\{\cS_{g}\}]=\sum_{g} \sum_{s \in \cS_{g}} \int_{\cM_{g}} [D\gamma] \int DX \prod_{\Cyl_{i}(\gamma) \in \cC^{2}(\gamma)}
 W^{(2)}[\Cyl_{i}(\gamma)],
 \label{2DFE}
\end{equation}
\end{widetext}
where $W^{(2)}$ and $\cS_{g}$ are  given by (\ref{HolCyl}) and (\ref{Sg2}), respectively.
In a general model, the pair $(G,B)$ for each cylinder in (\ref{2DFE}) can be taken as the sum of irreducible parts with coupling constants $\lambda_{i}^{(r)}$ and $\tilde{\lambda}_{i}^{(r)}$:
\begin{eqnarray}
 G_{i}=\sum_{r} \, \lambda_{i}^{(r)}\,G^{(r)},\,\,
 B_{i}=\sum_{r} \, \tilde{\lambda}_{i}^{(r)}\,B^{(r)},
 \label{GDoubl}
 \end{eqnarray}
where $i$ and $r$ are indices for the cylinders and bundles, respectively.
The extended moduli space (\ref{ExMod2}) with (\ref{DiffInv2}) is valid for the model (\ref{2DFE}) but we do not permute the cylinders in (\ref{ConDI2})
with the different gauge doublets (\ref{GDoubl}).
We call the model (\ref{2DFE}) special 2-gauge strings.

Let us discuss important configurations within the construction of 2-gauge strings:\\
\noindent
$\bullet$ {\bf Closed or open surfaces with one or two discs}\\
As an example of Eq.~(\ref{2DFE}), we add one or two discs on a surface $\Sigma$:
\begin{eqnarray}
 &I:&\,(\Sigma,\,\, \Disc_{1}^{(+)},\, \Disc_{2}), \label{DiscSurf1}\\
 \label{DiscSurf2}
 &II:&\,(\Sigma,\,\, \Disc_{1}^{(+)}).
 \end{eqnarray}
The discs in (\ref{DiscSurf1}) and (\ref{DiscSurf2}) cover the whole surface $\Sigma$.
The structure function $\cS_{g}$ is given simply by $[\bar{\gamma},(\Disc_{1}^{(+)}\,\{,\Disc_{2}\})]$ for each $g \in {\rm CTop}_{2}^{+}$ and $\bar{\gamma}$ is a section
of the bundle $\Met(\Sigma)\rightarrow\cM(\Sigma)$.
The sign $(+)$ in (\ref{DiscSurf1}) and (\ref{DiscSurf2}) means the same orientation compared with the orientation of the surface.
The partition function (\ref{2DFE}) for the configurations  (\ref{DiscSurf1}) and (\ref{DiscSurf2}) reads
\begin{widetext}
\begin{eqnarray}
 &I:&\,\,W_{F}^{(2)}[\{\cS_{g}\}]=\sum_{g} \int_{\cM_{g}} [D\gamma] \int DX\,       
 W^{(2)}[\Disc_{1}^{(+)}]_{S}\,W^{(2)}[\Disc_{2}]_{NA}\label{DInt1},\\\label{DInt2}
 &II:&\,\,W_{F}^{(2)}[\{\cS_{g}\}]=\sum_{g} \int_{\cM_{g}} [D\gamma] \int DX\, W^{(2)}[\Disc_{1}^{(+)}]_{S+NA},
\end{eqnarray}
\end{widetext}
here $S$ denotes the pair $(G,B)$ in the singlet representation, i.e., proportional to the unit matrix and $NA$ is the non-Abelian part.\\
%This decomposition together with the relation (\ref{ConDI2}) can serve as a definition of non-abelian Wilson surface.\\
\noindent
$\bullet$ {\bf Open stringlike configuration}\\
Let an open string surface $\Sigma$ be composed in the form
\begin{widetext}
\begin{eqnarray}
(\Sigma,\, \Disc^{(+)},\,\Cyl_{1},\ldots,\Cyl_{n-1}),\,\, (\Disc^{(+)},\,\Cyl_{1},\ldots,\Cyl_{n-1})\in (\cL\cP(\Sigma))^{n},
\label{OpSurfCom}
\end{eqnarray}
\end{widetext}
where $n-1$ is the number of holes.
In (\ref{OpSurfCom}) the disc and $\bar{A}$-boundaries (\ref{CylAB}) of the cylinders cover the whole surface and its boundaries, respectively.
Let $\bar{\gamma}$ be a section of the bundle $\Met(\Sigma)\rightarrow \cM(\Sigma)$. The structure function is taken  $[\bar{\gamma},(\Disc^{(+)},\Cyl_{1},\ldots,\Cyl_{n-1})]$.
The partition function (\ref{2DFE}) for the open configuration is
\begin{widetext}
\begin{eqnarray}
 W_{F}^{(2)}[\{\cS_{g}\}]=\sum_{g} \int_{\cM_{g}} [D\gamma] \int DX\,
 W^{(2)}[\Disc^{(+)}]_{S}\,\prod_{i=1}^{n-1}\,W^{(2)}[\Cyl_{i}]_{NA},
 \label{OpenCyl}
\end{eqnarray}
\end{widetext}
where the cylinders induce non-Abelian interaction.

The insertion of the $\bar{A}$-holonomy (\ref{HAcyl}) in Eq.~(\ref{HolCyl}) yields
\begin{widetext}
\begin{eqnarray}
 W^{(2)}[\Cyl]=
 \Tr \left \{\Hol(\bar{A},1)\,P_{s} ~\ex \left[-\frac{1}{2} \int_{0}^{1} ds\, \Hol(\bar{A},s)^{-1}\, SH(G,B,A,s)\,\Hol(\bar{A},s)\right]\right \}.
 \label{HCylMod}
\end{eqnarray}  
\end{widetext}
Using (\ref{HCylMod}) in Eq.~(\ref{OpenCyl}) we get the open string partition function  for the configuration when the cylinders fields $(G,B)$ are zero.
A model with moving Wilson loops is recovered if we exchange for the cylinder the $\bar{A}$-boundary (\ref{CylAB}) and the free boundary
\begin{equation}
 \Cyl(\bar{\bullet},1),\,\bar{\bullet}=[0,1].
 \label{CylA1}
\end{equation} 
$\bullet$
{\bf Non-Abelian branched space-time}\\
The singlet part in (\ref{DInt1}) and (\ref{DInt2}) stabilizes space-time and the non-Abelian part causes a branching of space-time. The Abelian and non-Abelian terms are separated by the new coupling constants.\\ 
\section {Conclusions}
In this paper, the new gauge interaction has been obtained for bosonic string. The path integral is equipped by moving one- and two-dimensional gauge loops.
The resulting partition function has acquired as a parameter the large moduli space dependence.

Straightforwardly, it is possible to generalize the special 2-gauge strings construction (\ref{2DFE}) by adding non-Abelian dilaton or tachyon terms.

As a consequence of the space-time non-Abelian structure, some nontraditional Higgs mechanism can be expected.
The model (\ref{DInt2}) corresponds  to an interacted parallel space-time. Such bigravity interaction in the field limit has been studied
in the literature \cite{Damour1} with cosmological applications.

The special connection on the space of horizontal path can be iterated into a higher connection on the multiple
horizontal path space \cite{CatST}. The higher connection is generated by a tower of non-Abelian $p-$forms ($p\geq1)$ and its Wilson loop can describe the non-Abelian interaction of a higher dimensional brane.

\acknowledgments
It is a pleasure to thank J. Horsk\'{y} for fruitful conversation and useful comments on the manuscript.
%\bibliography{Estring}% Produces the bibliography via BibTeX.

\end{document}